\documentstyle[12pt]{article}
\newcommand{\be}{\begin{equation}}
\newcommand{\ee}{\end{equation}}
\newcommand{\ba}{\begin{eqnarray}}
\newcommand{\ea}{\end{eqnarray}}

\begin{document}
\begin{center}
{\bf THREE-DIMENSIONAL SOLUTIONS OF SUPERSYMMETRICAL INTERTWINING RELATIONS
AND PAIRS OF ISOSPECTRAL HAMILTONIANS}\\
\vspace{0.5cm} {\large \bf F. Cannata$^{1,}$\footnote{E-mail:
cannata@bo.infn.it}, M.V. Ioffe$^{2,}$\footnote{E-mail:
m.ioffe@pobox.spbu.ru} ,
D.N. Nishnianidze$^{2,3,}$\footnote{E-mail: cutaisi@yahoo.com}}\\
\vspace{0.2cm}
$^1$ INFN, Via Irnerio 46, 40126 Bologna, Italy.\\
$^2$ Department of Theoretical Physics, Sankt-Petersburg State University,\\
198504 Sankt-Petersburg, Russia\\
$^3$ Akaki Tsereteli State University, 4600 Kutaisi, Georgia
\end{center}
\vspace{0.2cm} \hspace*{0.5in}
\hspace*{0.5in}
\begin{minipage}{5.0in}
{\small The general solution of SUSY intertwining relations for
three-dimensional Schr\"odinger operators is built using the class
of second order supercharges with nondegenerate constant metric. This solution
includes several models with arbitrary parameters. We are
interested only in quantum systems which are not amenable to
separation of variables, i.e. can not be reduced to lower
dimensional problems. All constructed Hamiltonians are partially
integrable - each of them commutes with a symmetry operator of
fourth order in momenta. The same models can be considered also
for complex values of
 parameters leading to a class of non-Hermitian isospectral Hamiltonians.}
\\
\vspace*{0.1cm}
\end{minipage}
\vspace*{0.2cm}
\section{\bf Introduction.}
\vspace*{0.1cm} \hspace*{3ex}
The method of supersymmetrical (SUSY) intertwining relations appeared naturally in the framework of a
supersymmetrical approach to nonrelativistic Quantum Mechanics \cite{susy}. Briefly speaking in the most general
formulation the method of SUSY intertwining relations for investigation of different models in Quantum Mechanics means
the following \cite{abei}. Let us try to solve the operator equations\footnote{From a pure mathematical point of view,
the problem can be formulated as follows. We look for two different factorizations
of the operator of fourth order in derivatives (\ref{intertw1}) such that $H_{1,2}$ are of Schr\"odinger form, and $Q^+$ is the
same intertwining second order multiplier.}
\ba
H_1 Q^+=Q^+H_2; \label{intertw1}
\ea
where $H_{1,2}$ is a pair of Schr\"odinger Hamiltonians, and the intertwining operator $Q^{+}$ is called the supercharge.
 This relation and its Hermitian conjugate with supercharge $Q^-=(Q^+)^{\dagger}$
lead to the isospectrality of Hamiltonians $H_{1,2}$ up to possible zero modes of supercharges.
The bound state eigenfunctions of $H_{1,2}$ are related (up to normalization factors) by the supercharges:
\ba
&&H_{1,2}\Psi^{(1),(2)}_n(\vec x)=E_n\Psi_n^{(1),(2)}(\vec x);\quad H_{1,2}=-\Delta^{(3)}+ V_{1,2}(\vec x); \label{schr}\\
&&\Psi_n^{(2)}(\vec x)=Q^-\Psi_n^{(1)}(\vec x);\quad \Psi_n^{(1)}(\vec x)=Q^+\Psi_n^{(2)}(\vec x);
\quad \vec x=(x_1,x_2,x_3); \quad n=0,1,2,...
\label{psi}
\ea
If $Q^+$ has some zero modes, and they coincide with the wave functions
of $H_2$ (or correspondingly, $Q^-$ has some zero modes coinciding with the wave functions
of $H_1$), these wave functions are annihilated according to (\ref{schr}) and have no analogous states in the spectra
 of the partner Hamiltonian.

This scheme works independently on the nature and specific properties of operators $H_{1,2}$.
In particular, it was realized in one-dimensional space with $Q^{\pm}$ of first order in derivatives both for scalar
 \cite{susy} and matrix \cite{matrix} potentials in the stationary formulation of Schr\"odinger equation.
The case of nonstationary Schr\"odinger equation was studied in \cite{junker}. The SUSY intertwining relations with
second order supercharges $Q^{\pm}$ were introduced first in one-dimensional case in \cite{second}, including new
irreducible supercharges which cannot be factorized with two first order multipliers.

A very useful role was played by second order supercharges in the case of two-dimensional SUSY intertwining relations
\cite{david}. A class of solutions of such intertwining relations was obtained. Each of these Hamiltonians is
completely integrable, commuting with its nontrivial symmetry operator of fourth order in momenta $R_1=Q^+Q^-$ or
$R_2=Q^-Q^+.$ Sometimes, besides being completely integrable the model is partially solvable: part of its spectrum
and corresponding wave functions are constructed by special methods like shape invariance and SUSY
 separation of variables \cite{new}, \cite{david}. In the case of two-dimensional generalization of Morse potential
 the exact solvability of the model was proven \cite{morse}. It is necessary to stress that only models, which do
 not allow for conventional separation of variables, were considered in the papers on two-dimensional Quantum
 Mechanics mentioned above.

Some attempts were made to generalize the method of SUSY intertwining relations to the physically most
interesting case of three-dimensional space. First, the direct generalization of formalism with the first order
supercharges was constructed for arbitrary dimensionality of space in \cite{abei}.
This approach has a specific property: the intertwining relations link scalar Hamiltonians with a chain of matrix
Schr\"odinger-like operators. These Hamiltonians can be interpreted as operators describing quantum systems with
 some internal degrees of freedom (for example, spin \cite{pauli}), but rather frequently this property seems to
  be inconvenient. Intertwining of two scalar Hamiltonians by first order supercharges was shown \cite{kuru} to
  offer only solutions with conventional separation of variables.
The only alternative generalization \cite{yamada}, to our knowledge, is analogous to the idea elaborated earlier
in two-dimensional case \cite{david}: to use the intertwining relations between pairs of scalar Hamiltonians but
with second order supercharges. In paper \cite{yamada} the particular solutions of such intertwining relations were
 found for the special form of second order supercharges
 $\frac{\partial^2}{\partial^2x_1}-\frac{\partial^2}{\partial^2x_2}+...$.

In the present paper we will consider the most general form of supercharges $Q^{\pm}$ with {\bf constant metric}
$g_{ik}$, i.e. $Q^{\pm}=g_{ik}\frac{\partial}{\partial x_i}\frac{\partial}{\partial x_k}+...,$
where the summation over repeated indices $i,k=1,2,3$ is implied here and below. In contrast to the similar approach
in two-dimensional space \cite{david}, where a list of particular solutions was found, in three-dimensional case
the general solution of intertwining relations will be built for arbitrary constant metric $g_{ik}$
excluding the degenerate (up to rotations) case of $g_{ik}=(1,0,0)$. This difference is due to a more restrictive
character of intertwing relations in higher dimensions. The general solution of three-dimensional intertwining
with constant metric in supercharge includes several classes of models, which depend on some parameters.
These parameters can be chosen either real or complex leading to Hermitian or non-Hermitian partner Hamiltonians.

\section{\bf General solution of second order intertwining relations with nondegenerate metric.}
\vspace*{0.1cm} \hspace*{3ex}
As announced in Introduction, we shall study the intertwining (\ref{intertw1}) with the most general supercharges of
second order in derivatives with constant metric (highest order coefficients) $g_{ik}$:
\be
Q^+=g_{ik}\partial_i\partial_k+C_i(\vec x)\partial_i+B(\vec x);\quad \partial_i\equiv \frac{\partial}{\partial x_i}.
\label{QQ}
\ee
By space rotations the matrix $g_{ik}$ can be diagonalized to $g_{ik}=g_{ii}\delta_{ik}.$
At first, we will be interested in situations when all three diagonal elements $g_{ii}$ after such rotation
do not vanish. The case $g_{ii}=(1,-1,0)$
was considered in \cite{yamada}, the case $g_{ii}=(1,1,0)$ will be considered later in this Section,
the case $g_{ii}=(1,a,0);\,\, a\neq \pm 1$ gives no interesting solutions for (\ref{intertw1}), and for the degenerate case
$g_{ii}=(1,0,0)$ only some particular solutions will be given in the very end of this Section.

Thus, let us consider now the supercharges $Q^{\pm},$ which by a suitable normalization can be reduced
to two classes of $g_{ik}$:
\ba
&&(A)\quad\quad g_{ii}=(1,1,d);\quad d\neq 0\,\, d\neq 1 \label{11d}\\
&&(B)\quad\quad g_{ii}=(1,d_2,d_3);\quad d_3\neq 0,1;\,\, d_2\neq 0,1;\,\, d_2\neq d_3 .\label{1dd}
\ea

The intertwining relations (\ref{intertw1}) for constant $g_{ik}$ can be rewritten as a system of $6+3+1=10$ differential
equations, some of them nonlinear:
\ba
&&\partial_iC_k(\vec x)+\partial_kC_i(\vec x)=2V(\vec x)g_{ik}; \label{D.2.1}\\
&&\Delta^{(3)}C_i(\vec x)+2\partial_iB(\vec x)+2g_{ik}\partial_kV_2=2V(\vec x)C_i(\vec x);
\label{D.2.2}\\
&&\Delta^{(3)}B(\vec x)+g_{ik}\partial_i\partial_kV_2(\vec x)+C_i(\vec x)\partial_iV_2(\vec x)=
2V(\vec x)B(\vec x); \label{D.2.3}\\
&&V_1(\vec x)-V_2(\vec x)\equiv 2V(\vec x). \label{V}
\ea

The six equations (\ref{D.2.1}) and the defining Eq.(\ref{V}) can be combined:
\ba
&&\partial_1C_1(\vec x)=d_2^{-1}\partial_2C_2(\vec x)=d_3^{-1}\partial_3C_3(\vec x)=V(\vec x);\label{1.1}\\
&&\partial_iC_k(\vec x)+\partial_kC_i(\vec x)=0.;\quad i\neq k. \label{1.2}
\ea
After further differentiation of (\ref{1.2}) in respect to $x_j$ with $j\neq i,\,j\neq k$ we obtain:
\ba
&&C_1(\vec x)=f_3(x_1,x_2)+f_2(x_1,x_3);\nonumber\\
&&C_2(\vec x)=[g_1(x_2,x_3)+g_3(x_1,x_2)]d_2;\nonumber\\
&&C_3(\vec x)=[h_1(x_2,x_3)+h_2(x_1,x_3)]d_3,\nonumber
\ea
where $f, g, h$ are auxiliary functions specified in the following.

By substitution of these relations into (\ref{1.1}), we get the functional-differential equations,
which are integrated in a general form:
\ba
&&g_1(x_2,x_3)=x_2G_3^{\prime}(x_3)+G_2(x_2)+K_3(x_3);\nonumber\\
&&f_2(x_2,x_3)=x_1G_3^{\prime}(x_3)+G_1(x_1)+M_3(x_3);\nonumber\\
&&f_3(x_1,x_2)=\tilde G_1(x_1)+x_1\tilde G_2^{\prime}(x_2)+K_2^{\prime}(x_2)-G_1(x_1);
\nonumber\\
&&g_3(x_1,x_2)=\tilde G_2(x_2)+x_2\tilde G_1^{\prime}(x_2)+K_1^{\prime}(x_1)-G_2(x_2);
\nonumber\\
&&h_2(x_1,x_3)=x_3\tilde G_1^{\prime}(x_1)+\tilde L_1(x_1)+M_3(x_3);\nonumber\\
&&h_1(x_2,x_3)=x_3\tilde G_2^{\prime}(x_2)+L_2(x_2)+G_3(x_3)-M_3(x_3).\nonumber
\ea
Therefore, we have the expressions for coefficient functions $C_i(\vec x)$ of the supercharges in terms of the eleven
arbitrary functions above $G_i(x_i), \tilde G_{1}(x_{1}), \tilde G_{2}(x_{2})... .$ If these expressions are inserted
into equations (\ref{1.2}),
the eleven functions above will be specified being polynomials maximally of second order in space coordinates.
Thus the general solution of (\ref{1.1}) takes the form:
\ba
C_1(\vec x)&=&a_1(x_1^2-d_2x_2^2-d_3x_3^2)+2a_2x_1x_2+2a_3x_1x_3+bx_1-b_1x_2-b_3x_3+\kappa_1;\nonumber\\
C_2(\vec x)&=&a_2(-x_1^2+d_2x_2^2-d_3x_3^2)+2a_1d_2x_1x_2+2a_3d_2x_2x_3+bd_2x_2+b_1x_1-b_2x_3+\kappa_2;
\label{4.1}\\
C_3(\vec x)&=&a_3(-x_1^2-d_2x_2^2+d_3x_3^2)+2a_1d_3x_1x_3+2a_2d_3x_2x_3+bd_3x_3+b_3x_1+b_2x_2+\kappa_3;
\nonumber
\ea
and the difference between potentials $V_{1,2}$ is given now by:
\be
V(\vec x)=2(a_1x_1+a_2x_2+a_3x_3)+b.
\nonumber
\ee

The next step consists in solving the three equations (\ref{D.2.2}), where the first Laplacian terms are constants
 defined from (\ref{4.1}):
\ba
&&\partial_1B(\vec x)+\partial_1V_2(\vec x)=V(\vec x)C_1(\vec x)-c_1;\quad c_1\equiv a_1(1-d_2-d_3);\nonumber\\
&&\partial_2B(\vec x)+d_2\partial_2V_2(\vec x)=V(\vec x)C_2(\vec x)-c_2;\quad c_2\equiv a_2(d_2-1-d_3);\label{4.4}\\
&&\partial_3B(\vec x)+d_3\partial_3V_2(\vec x)=V(\vec x)C_3(\vec x)-c_3;\quad c_3\equiv a_3(d_3-1-d_2).\nonumber
\ea

Now we will restrict ourselves to the case of metric (\ref{11d}), i.e. $d_2=1,\,d_3\equiv d\neq 0,1.$ Then, after the
simple manipulations (derivatives and linear combinations) with (\ref{4.4}) we obtain the necessary conditions:
\begin{equation}\label{5.0}
a_1=a_2=b_1=0.
\end{equation}
As for the constant $a_3,$ it is convenient to consider separately
two options: $(i)\,\, a_3\neq 0$ and $(ii)\,\, a_3 = 0.$

In the case $(i)\,\, a_3\neq 0$ a suitable translation of space coordinates allows to cancel the linear terms in
(\ref{4.1}):
\begin{equation}\label{5.1}
C_1(\vec x)=2a_3x_1x_3+e_1;\, C_2(\vec x)=2a_3x_2x_3+e_2;\, C_3(\vec x)=a_3(dx_3^2-x_2^2-x_1^2)+e_3;\, V(\vec x)=2a_3x_3,
\end{equation}
where $e_i$ are new arbitrary constants. Taking into account the numerical values of the constants, Eq.(\ref{5.0}),
we obtain for the considered case
$(d_2=1;\,\,a_3\neq 0)$ the general formulas:
\ba
V_2(\vec x)&=&\frac{a_3^2d}{2(d-1)}x_3^4-\frac{1}{d-1}[3a_3^2x_3^2(x_1^2+x_2^2)-a_3e_3x_3^2+c_3x_3+\nonumber\\
&+& 2a_3x_3(e_1x_1+e_2x_2)+q_3(x_3)-F_3(x_1,x_2)];
\label{6.1}
\ea
\ba
B(\vec x)&=&-\frac{a_3^2d}{2(d-1)}x_3^4+\frac{1}{d-1}[(2d+1)a_3^2x_3^2(x_1^2+x_2^2)-a_3e_3x_3^2+c_3x_3+
\nonumber\\
&+&2da_3x_3(e_1x_1+e_2x_2)++dq_3(x_3)-F_3(x_1,x_2)],
\label{6.2}
\ea
where $q_3(x_3)$ and $F_3(x_1,x_2)$ are arbitrary functions, which will be defined from the last intertwining relations
 Eq.(\ref{D.2.3}).

We use the divergence of vector equation (\ref{D.2.2}) to transform (\ref{D.2.3}) to:
\begin{equation}\label{6.3}
-(1+\frac{d}{2})\Delta^{(3)}V(\vec x)+C_i(\vec x)\partial_iV(\vec x)+(2+d)V^2(\vec x)+C_i(\vec x)\partial_iV_2(\vec x)-
2V(\vec x)B(\vec x)=0.
\end{equation}
Since the l.h.s. can be written as a polynomial in $x_3,$ we put
the coefficients to zero. Inserting explicit expressions
(\ref{5.1}), (\ref{6.1}), (\ref{6.2}) into (\ref{6.3}) and after
rather long but straightforward algebra, we derive that the
solution exists only for the metric $(1,1,-1),$ i.e. for $d=-1.$
In this case the function $q_3$ is linear (with arbitrary constant
$\alpha_0$):
\begin{equation}
q_3(x_3)=-a_3x_3+\alpha_0 . \nonumber
\end{equation}
As for the function $F_3,$ it is defined by two differential equations:
\begin{equation}\label{7.2}
-2(F_3(x_1,x_2)+\alpha_0)-(x_1\partial_1+x_2\partial_2)F_3(x_1,x_2)+(e_1^2+e_2^2)+(e_3-a_3\rho^2)(3a_3\rho^2-e_3)=0;
\end{equation}
\begin{equation}\label{7.3}
(e_1\partial_1+e_2\partial_2)F_3(x_1,x_2)=2a_3(e_3-a_3\rho^2)(e_1x_1+e_2x_2),
\end{equation}
where we introduced the radial coordinate in the plane $\rho^2\equiv x_1^2+x_2^2.$

In the particular case of constants $e_1=e_2=0$ the equation (\ref{7.3}) is trivially satisfied, and the general solution
for $F_3$ can be represented in polar coordinates $\rho ,\,\phi $ as:
\begin{equation}
F_3(x_1,x_2)=-\frac{1}{2}a_3^2\rho^4+a_3e_3\rho^2-\frac{2f(\phi)}{\rho^2}-\frac{1}{2}(2\alpha_0+e_3^2).\nonumber
\end{equation}
Thus we obtain the expressions for the partner potentials:
\begin{equation}\label{8.1}
V_{1,2}(\vec x)=\frac{a_3^2}{4}(x_3^4+\rho^4)+\frac{3a_3^2}{2}x_3^2\rho^2-\frac{a_3e_3}{2}(x_3^2+\rho^2)\pm 2a_3x_3+
\frac{f(\phi)}{\rho^2}+Const,
\end{equation}
which are anharmonic oscillators of fourth order with additional $1/\rho^2$ term.

In the particular case $e_1e_2\neq 0$ it is convenient to introduce the new space coordinates:
\begin{equation}
y_1=\frac{x_1}{e_1}+\frac{x_2}{e_2};\,\,y_2=\frac{x_1}{e_1}-\frac{x_2}{e_2},\nonumber
\end{equation}
so that in (\ref{7.2}), (\ref{7.3})
\begin{equation}
e_1\partial_1+e_2\partial_2=2\partial_{y_1};\quad x_1\partial_1+x_2\partial_2=y_1\partial_{y_1}+y_2\partial_{y_2}.
\nonumber
\end{equation}
Then the equation (\ref{7.3}) can be solved in a general form:
\ba
&&F_3(x_1,x_2)=-\frac{a_3^2(e_1^2+e_2^2)^2}{32}y_1^4-\frac{a_3^2(e_1^4-e_2^4)}{8}y_1y_2(y_1^2+y_2^2)
-\nonumber\\
&&-\frac{a_3^2(3e_1^4+3e_2^4-2e_1^2e_2^2)}{16}y_1^2y_2^2+
\frac{a_3e_3}{4}[(e_1^2+e_2^2)y_1^2+2(e_1^2-e_2^2)y_1y_2]+p_2(y_2),
\label{8.4} \ea with a function $p_2(y_2)$ defined from
(\ref{7.2}):
\begin{equation}
p_2(y_2)=\frac{a_3e_3(e_1^2+e_2^2)}{4}y_2^2-\frac{e_3^2-e_1^2-e_3^2+2\alpha_0}{2}+\frac{\kappa}{y_2^2}-
\frac{a_3^2(e_1^2+e_2^2)^2}{32}y_2^4.\nonumber
\end{equation}
Now, inserting this function into (\ref{8.4}) and going back to the
variables $x_i,$ we obtain the final expression for potentials
$V_{1,2}:$
 \ba
 &&V_{1,2}(\vec x)=\frac{a_3^2}{4}(x_3^4+\rho^4)+\frac{3a_3^2}{2}x_3^2\rho^2-\frac{a_3e_3}{2}(x_3^2+\rho^2)+
 a_3e_3(e_1x_1+e_2x_2)-\nonumber\\ &&-\frac{\kappa e_1^2e_2^2}{2(e_2x_1-e_1x_2)^2}\pm 2a_3x_3+Const.\label{9.2}
 \ea

The last option for the values of $e_1,e_2$ we have to consider is the case with one of them vanishing and the
other not. For definiteness, let us take $e_2=0.$ Then the general solution of (\ref{7.3}) is:
\begin{equation}
F_3(x_1,x_2)=\frac{a_3}{2}x_1^2(2e_3-a_3x_1^2-2a_3x_2^2)+q_2(x_2), \nonumber
\end{equation}
with an arbitrary function $q_2(x_2).$ Using it in (\ref{7.2}), we obtain a first order differential equation
for $q_2(x_2),$ which admits the general solution:
\begin{equation}
q_2(x_2)=-\frac{1}{2}a_3^2x_2^4+a_3e_3x_2^2+\frac{2\mu}{x_2^2}+\frac{1}{2}(e_1^2-e_3^2-2\alpha_0) \nonumber
\end{equation}
($\mu$ is an arbitrary integration constant). The resulting potential has the form:
\begin{equation}\label{10.3}
V_{1,2}(\vec x)=\frac{a_3^2}{4}(x_3^4+\rho^4)+\frac{3a_3^2}{2}x_3^2\rho^2-
\frac{a_3e_3}{2}(x_3^2+\rho^2)+a_3e_1x_1x_3\pm 2a_3x_3-
\frac{\mu}{x_2^2}+Const
\end{equation}

In the case $(ii)\, a_3=0$ equations (\ref{4.1}), (\ref{5.0})
together with a suitable shift of space coordinates allow to
conclude that the coefficient functions $C_i(\vec x)$ become
linear functions:
 \be C_1(\vec x)=bx_1-b_3x_3;\,C_2(\vec x)=bx_2-b_2x_3;\,C_3(\vec x)=bdx_3+b_3x_1+b_2x_2;\,
 V(\vec x)=b \label{11.1}
 \ee
with arbitrary real constants $b,b_2,b_3.$ Then from the explicit
formulas for $V_{1,2}$ we find that both partner Hamiltonians
$H_{1,2}$ being nonisotropic harmonic oscillator are amenable to
separation of variables.

But there is just one solution for the case $a_3=0$ which does not
allow separation of variables. In order to obtain it, we have to
extend the original assumption about the constants in (\ref{11.1}). Indeed,
if we allow $b_2$ and $b_3$ to be complex numbers with $b_2=ib_3,$ after simple calculations
we obtain:
 \ba &&V_2(\vec x)=\frac{b^2}{4}(x_1^2+x_2^2+x_3^2)-bb_3(x_1+ix_2)x_3+
 \frac{b_3^2}{4}(x_1+ix_2)^2 + \nonumber\\&&
 +\frac{n}{(x_1+ix_2)^2}+\omega (x_1+ix_2)+Const,\label{D.32.1}
 \ea
 \be V_1(\vec x)=V_2(\vec x)+2b, \label{D.32.22}
 \ee
with arbitrary constants $n,\omega.$

Both potentials are second order polynomials with complex
coefficients\footnote{The complex coefficients appear in the
supercharges $Q^{\pm}$ also. See detail discussion of one-dimensional case in
\cite{sokolov} and of two-dimensional case in \cite{complex2}.} and they differ from each other by the
arbitrary
constant $b.$ Such pair of intertwined Hamiltonians realize the
simplest kind of shape invariance \cite{shape}, \cite{new}. In this context,
it means that if the Hamiltonian $H_1$ has the wave function
$\Psi_k^{(1)}$ with eigenvalue $E_k$ (see Eq.(\ref{schr})), then
the function $\Psi_k^{(2)}=Q^-\Psi_k^{(1)}$ is also wave function
of the same Hamiltonian $H_1$ with eigenvalue $E_k+2b.$ Therefore,
an arbitrary eigenvalue $E_0$ in the spectra of $H_1$ is
accompanied by the tower of eigenstates with energies $E_0+2bk$.
The same is obviously true for $H_2.$ We must stress that this
property of the spectra are formulated up to possible zero modes
of supercharges $Q^{\pm}.$ These zero modes can, in principle,
truncate the tower described above.

An additional property of the spectrum of model (\ref{D.32.1}) is provided by its symmetry.
Indeed, for real values of parameter $n$ both Hamiltonians $H_{1,2}$ are invariant under the
combined reflection $P_2,$ i.e. $x_2\to -x_2,$ and time reflection $T.$ Due to this $P_2T$-symmetry,
the spectrum of Hamiltonians consists of \cite{PT1}, \cite{PT2} real eigenvalues $E=E^{\star}$
 and complex conjugate pairs $E, \, E^{\star}.$

One can notice that all models above which were supposed to have real potentials can be easily
complexified also by chosing complex values of parameters in (\ref{8.1}), (\ref{9.2}), (\ref{10.3}).
Although no property of shape invariance is observed for these potentials, some invariances
under combined reflections can be realized with suitable choice of complex parameters. For example,
$P_1T-$symmetry of potential (\ref{9.2}) for pure imaginary $e_1$ and other parameters real.

Let us consider now the case when the metric can be reduced to
\be
(C)\quad\quad g_{ik}=(1,1,0). \quad\quad\quad\nonumber
\ee
Then the general solution of Eq.(\ref{D.2.1}) is:
 \ba
 &&C_1=-x_3(2\alpha x_1+\alpha_1)-2(\nu x_1+\nu_1);\,C_2=-x_3(2\alpha x_2+\alpha_2)-2(\nu
 x_2+\nu_2);\nonumber\\ &&C_3=\alpha
 (x_1^2+x_2^2)+\alpha_1x_1+\alpha_2x_2+\beta;\, V=-2(\alpha
 x_3+\nu),\label{1}
 \ea
 with arbitrary constants $\alpha , \alpha_{1,2}, \nu , \nu_{1,2}.$
For $\alpha\neq 0,$ it can be transformed by translations to:
 \ba
 C_1=-2(\alpha x_1x_3+\nu_1),\, C_1=-2(\alpha x_2x_3+\nu_2); \, C_3=\alpha (x_1^2+x_2^2)+\beta; \,
 V=-2\alpha x_3.\nonumber
 \ea
The first two equations of Eq.(\ref{D.2.2}) lead to:
 \be
 B+V_2=2\alpha x_3[\alpha
 (x_1^2+x_2^2)x_3+2(\nu_1x_1+\nu_2x_2)]+f_3(x_3),\nonumber
 \ee
with an arbitrary function $f_3(x_3).$
The last equation of Eq.(\ref{D.2.2}) gives $B(\vec x)$:
 \be
 B(\vec x)=-\alpha x_3^2[\alpha (x_1^2+x_2^2)+\beta ]-2\alpha x_3+F_3(x_1,x_2), \nonumber
 \ee
where $F_3(x_1,x_2)$ is also an arbitrary function, and $\beta$ is a constant. Therefore the potential
$V_2(\vec x)$ is obtained:
 \ba
 V_2(\vec x)=3\alpha^2x_3^2\rho^2+\alpha\beta
 x_3^2+2\alpha (2\nu_1x_1+2\nu_2x_2+1)x_3+f_3-F_3.\nonumber
 \ea

The arbitrary functions $f_3(x_3), F_3(x_1,x_2)$ are defined by
Eq.(\ref{D.2.3}) using also Eq.(\ref{D.2.2}):
\ba
f_3(x_3)&=&4\alpha^2x_3^4+\beta_3x_3^3+\beta_2x_3^2+\beta_1x_3+\beta_0;\nonumber\\
F_3(x_1,x_2)&=&-\frac{1}{2}\alpha^2\rho^4-\frac{1}{2}\beta_2\rho^2-\frac{f(\phi)}{\rho^2},\nonumber
\ea
with arbitrary periodical function $f(\phi)$ of polar coordinate $\phi .$
The constants above must be chosen vanishing: $\nu_1=\nu_2=\beta_1=\beta_3=0$.
Finally, the partner potentials $V_{1,2}(\vec x)$ are obtained as:
 \ba
 V_{1,2}=\alpha^2(3x_3^2\rho^2+4x_3^4+\frac{1}{2}\rho^4)+\beta_2(x_3^2+\frac{1}{2}\rho^2)+
 \frac{f(\phi)}{\rho^2}\mp 2\alpha x_3.\label{8}
 \ea

In the case $\alpha =0,$ Eq.(\ref{1}) gives after translation:
 \ba
 C_1=-(\alpha_1 x_3+2\nu_1 x_1); \, C_2=-(\alpha_2 x_3+2\nu x_2); \,
 C_3=\alpha_1x_1+\alpha_2x_2; \, V=-2\nu ,\label{9}
 \ea
 and, using (\ref{D.2.2}):
 \ba
 &&B=-2\nu(\alpha_1x_1+\alpha_2x_2)x_3+N_3(x_1,x_2);\label{10}\\
 &&V_2=2\nu^2\rho^2+4\nu(\alpha_1x_1+\alpha_2x_2)+n_3(x_3)-N_3(x_1,x_2),\label{11}
 \ea
with arbitrary functions $n_3(x_3),N_3(x_1,x_2).$
The analysis of Eqs.(\ref{9})-(\ref{11}) together with Eqs.(\ref{D.2.2}), (\ref{D.2.3})
shows that the only solution with real parameters corresponds to models allowing standard separation
of variables, which are not studied in this paper.

The only constant metric $g_{ik}$ in $Q^{\pm}$, which does not allow to find the general solution, is
(up to rotations) the degenerate one - $g_{ii}=(1,0,0).$ In this case only particular solutions were found:
\ba
 &&V_2=\frac{1}{9}(\alpha x_2+x_3)^2x_1^{-2/3}-\frac{5c}{3}(\alpha
 x_2+x_3)^2x_1^{2/3}+\frac{16c^2}{9(1+\alpha^2)}(\alpha
 x_2+x_3)^2-\nonumber\\&&-\frac{a_4}{1+\alpha^2}(\alpha x_2+x_3)
 +\frac{c^2}{4}x_1^2+\frac{1+\alpha^2}{2}x_1^{4/3}+\frac{3a_4}{8c}x_1^{2/3}+
 \frac{7}{36}x_1^{-2}+\Phi(\alpha x_3-x_2),\nonumber\\
 &&V=\frac{2}{9}(\alpha x_2+x_3)x_1^{-4/3}+c \nonumber
 \ea
 and
 \ba
 &&V_2=x_1^2[3(\alpha x_2+x_3)^2-3b(\alpha
 x_2+x_3)-2(b^2+c)]+\frac{1+\alpha^2}{2}x_1^4-\omega
 x_1^{-2}+\nonumber\\&&
 \frac{1}{1+\alpha^2}[4(\alpha x_2+x_3)^4-8b(\alpha x_2+x_3)^3-c(\alpha x_2+x_3)^2+
 \nonumber\\&&+b(4b^2+c)(\alpha x_2+x_3)]+\Phi(\alpha
 x_3-x_2),\nonumber\\&&
 V=-2(\alpha x_2+x_3)+b, \nonumber
 \ea
where $\Phi$ is an arbitrary function of its argument.

\section{\bf Conclusions.}
\vspace*{0.1cm} \hspace*{3ex}
The aim of the present paper was to provide an exhaustive analysis of intertwining relations for
 second order supercharges with constant metric in three-dimensional case. In this space the
 intertwining relations are rather restrictive: only specific classes of solutions exist.
 Namely, four different classes of real potentials were built in the framework of general solution
 of intertwining relations with nondegenerate metric - (\ref{8.1}), (\ref{9.2}), (\ref{10.3}), (\ref{8}).
 All of them have the form of anharmonic oscillator of fourth order in coordinates plus the term with $1/x^2-$dependence.
 In addition, particular solutions were obtained for degenerate metric $(1,0,0).$
 For all these models pairs of almost (up to zero modes of $Q^{\pm}$) isospectral Hamiltonians were obtained. Each model
 is at least partially integrable, since the Hamiltonians commute with the corresponding symmetry operators of
 fourth order in momenta.

Although the intertwining relations themselves do not provide, in general, the wave functions and energy eigenvalues,
in two-dimensional space the specific SUSY methods
- SUSY separation of variables and shape invariance \cite{new}, \cite{david}, \cite{morse} - allowed to find a
part or even the whole spectrum. In contrast to this situation, it seems to be impossible to apply these methods
in three-dimensional space: there is no shape invariance property and there are no tools to find zero modes of
$Q^{\pm}$ explicitly.

It was also useful to consider the obtained results from the point of view of Quantum Mechanics with non-Hermitian
Hamiltonians \cite{sokolov}, \cite{complex2}, \cite{PT1}, \cite{PT2}. In this case, the potentials with complex
values of parameters obey some discrete symmetries
with antilinear operators, leading to corresponding specific properties of complex spectra.

\section*{\bf Acknowledgments}
The work was partially supported by INFN and University of Bologna (M.V.I. and D.N.N.), by Russian grant
RNP 2.1.1/1575 (M.V.I.) and grant ATSU/09317 (D.N.N.).  
\end{document}